# Transport or Store? Synthesizing Flow-based Microfluidic Biochips using Distributed Channel Storage

Chunfeng Liu[1,2], Bing Li[1], Hailong Yao[3], Paul Pop[4], Tsung-Yi Ho[2,5], Ulf Schlichtmann[1]
[1]Institute for Electronic Design Automation, [2]Institute for Advanced Study, Technical University of Munich, Germany
[3]Tsinghua University, Beijing, China, [4]Technical University of Denmark, [5]National Tsing Hua University, Hsinchu, Taiwan
{chunfeng.liu, b.li, ulf.schlichtmann}@tum.de, hailongyao@tsinghua.edu.cn, paupo@dtu.dk, tyho@cs.nthu.edu.tw

## ABSTRACT

Flow-based microfluidic biochips have attracted much attention in the EDA community due to their miniaturized size and execution efficiency. Previous research, however, still follows the traditional computing model with a dedicated storage unit, which actually becomes a bottleneck of the performance of biochips. In this paper, we propose the first architectural synthesis framework considering distributed storage constructed temporarily from transportation channels to cache fluid samples. Since distributed storage can be accessed more efficiently than a dedicated storage unit and channels can switch between the roles of transportation and storage easily, biochips with this distributed computing architecture can achieve a higher execution efficiency even with fewer resources. Experimental results confirm that the execution efficiency of a bioassay can be improved by up to 28% while the number of valves in the biochip can be reduced effectively.

## 1 Introduction

Microfluidic biochips have reshaped the traditional biochemical experiment flow with their high execution efficiency and miniaturized fluid manipulation [1, 2]. With this miniaturization, biochemical assays can be scaled down to nanoliter volumes, so that precious reagents can be saved to reduce experiment cost. Since operations executed on a biochip are automated and controlled by a microcontroller, the reliability in executing biochemical experiments can also be improved significantly compared with the traditional manual experiment flow.

Microfluidic biochips based on continuous flow use valves to control the movement of samples and reagents. The structure of a valve is shown in Fig. 1(a). On a substrate, a flow channel is constructed for the transportation of fluids. Above the flow channel, a control channel connected to an air pressure source is used to control the open/closed state of the valve. Both channels are built from elastic materials, so that air pressure in the control channel squeezes the flow channel below to block the movement of the fluid. Conversely, if the pressure in the control channel is released, the fluid can resume its movement.

With valves as the basic controlling components, complex devices can be constructed. For example, mixers can be built from channels and valves to execute mixing operations, which are very common in biochemical assays. The structure of a mixer is shown in Fig. 1(b), where the three valves at the top are actuated alternately by applying and releasing air pressure in the control channels to form a circular flow around the device.

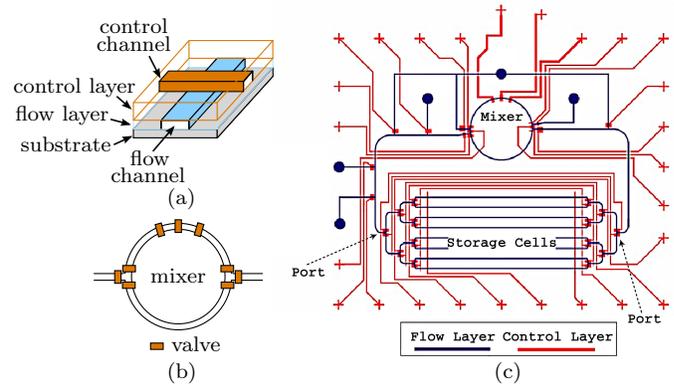

Figure 1: Components and structure of flow-based biochips. (a) Valve structure. (b) Mixer. (c) Biochip with eight storage cells [3].

The other six valves are used to control the entering and exiting of fluids.

Besides mixers, a dedicated storage unit can also be built from channels and valves. Fig. 1(c) shows a schematic of a biochip with a mixer connected to a storage unit. This storage unit contains eight side-by-side cells to store fluid samples. At a port of the storage unit, valves in a multiplexer-like structure direct a fluid sample to enter a specific cell or to leave the storage unit.

Biochips are used to execute operations in biochemical assays, whose protocols are usually described by sequencing graphs. In Fig. 2(a), the sequencing graph of the mixing phase of the polymerase chain reaction (PCR) is illustrated. This assay takes eight input samples ($i_1 \sim i_8$) and processes them with seven mixing operations ($o_1 \sim o_7$) to generate copies of a DNA sequence. The edges in the sequencing graph define the dependency of operations, where a parent operation should always be executed before its child operations. If for each operation a mixer is assigned, seven mixers have to be built on the chip. However, it is unusual to assign resource so freely due to cost. Instead, mixers are reused to execute the operations while maintaining their dependency as specified by the sequencing graph.

In recent years, design automation tools have been introduced to deal with design challenges of biochips. The method in [4] proposes a top-down flow to generate a biochip architecture while minimizing the execution time of the assay. The method in [5] proposes to solve the flow channel routing problem considering obstacles with an algorithm based on rectilinear Steiner minimum tree, while placement/routing iterations are performed in [6] to reduce flow-channel crossings. Control logic synthesis is investigated in [7] to reduce the number of control pins, in [8] to optimize valve switching considering the relation between control patterns, and in [9] to match lengths of control channels. In [10], flow layer and control layer codesign is proposed to achieve valid routing on both layers iteratively, and in [11] interactions on both flow and control layers are modeled by an ILP formula-tion to achieve a better codesign efficiency. In [12], fluid storage is investigated during synthesis with an ideal chip architecture. Furthermore, dynamic construction of devices and flow channels on a fully programmable valve array is explored in [13]. Fault models and test of manufacturing defects for flow-based biochips are discussed in [14, 15].

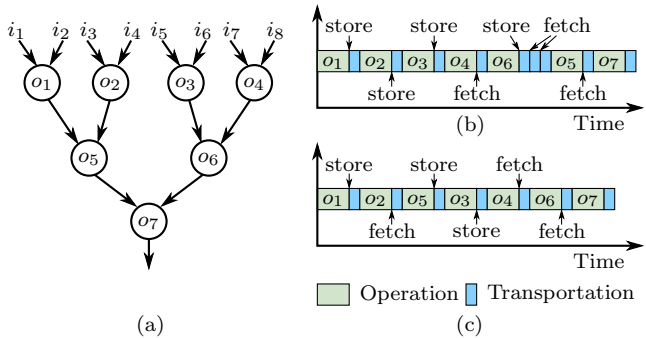

Figure 2: Sequencing graph and different schemes of scheduling with one mixer. (a) Sequencing graph of PCR. (b) Scheduling with four store operations. Required storage capacity is three. (c) Scheduling with three store operations. Required storage capacity is two. The execution time of the assay with the second schedule is shorter.

In this paper, we propose the first synthesis framework that models distributed channel storage and time multiplexing at intersections of flow channels, so that an assay can be executed very efficiently. Our contributions include:

- Instead of using a dedicated storage unit, we cache intermediate fluid samples in channels, so that not only the access bandwidth limit at the ports of the dedicated storage unit is overcome, but also the efficiency of channels and valves is improved.
- Time multiplexing at the intersections of channels is modeled considering all transportation tasks for the first time. With this model, transportation conflict can be avoided during architectural synthesis so that operations do not need to be postponed as in other methods.
- This is the first work to consider storage minimization from scheduling to architectural synthesis, so that the execution time of the assay can be reduced effectively.
- A compact physical design can be generated from the result of architectural synthesis easily. This design is already planar due to the direct modeling of switches at intersections of transportation channels.

The rest of this paper is organized as follows. In Section 2, we explain the motivation of storage synthesis and formulate the problem we address in this paper. In Section 3, we describe the techniques to reduce storage usage and to model distributed channel storage in scheduling and architectural synthesis. We also demonstrate how a compact physical design can be generated from the synthesized planar connection graph. Experimental results are reported in Section 4. Conclusions are stated in Section 5.

## 2 Motivation and Problem Formulation

The sequencing graph of an assay defines the dependency of operations. These operations are scheduled to devices in a given order for execution. Different schedules, however, yield different storage usage and transportation requirements. In Fig. 2(b) and 2(c), two schedules for the PCR assay are shown, where one mixer is used to execute the operations. The first schedule executes the operations in the order $o_1 \to o_2 \to o_3 \to o_4 \to o_6 \to o_5 \to o_7$. After executing $o_1$, the intermediate result should be transported to the storage unit, so that the device can be reused to execute $o_2$. When $o_6$ is executed, it takes the result of $o_4$ from the mixer directly as one input and fetches the result of $o_3$ from the storage unit. In this schedule, in total four storage operations and four fetch operations are needed. In addition, the results of $o_1$, $o_2$ and $o_3$ stay in the storage unit at the same time, so that the capacity requirement of the storage unit is three. In the schedule in Fig. 2(c) with the execution order $o_1 \to o_2 \to o_5 \to o_3 \to o_4 \to o_6 \to o_7$, however, there are only three store and three fetch operations, leading to a storage capacity of only two units. In addition, the execution time of the assay in the second schedule is even shorter.

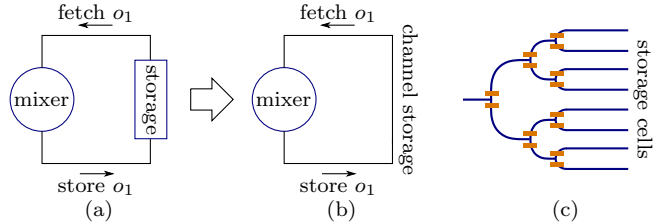

Figure 3: Storage mechanisms. (a) Storage with dedicated unit. (b) Channel storage. (c) Port of dedicated storage unit.

The comparison of the two schedules in Fig. 2(b) and 2(c) demonstrates that the schedule scheme affects the transportation of fluid samples as well as the required capacity of the storage unit, and the execution time of the assay may be unnecessarily prolonged if storage and transportation are not considered. This important problem, however, has not been dealt with by previous methods.

When a fluid sample is stored, it is transported to the storage unit through a channel. The diagram of a simple chip with one mixer and one storage unit is shown in Fig. 3(a). When this chip is used to execute the operation $o_1 \to o_2 \to o_5$ in the schedule in Fig. 2(c), the mixer first stores the result of $o_1$ in the storage unit. After $o_2$ is finished, the result of $o_1$ is fetched back to the mixer to execute the operation $o_5$. In this case, the channel itself is sufficient to store the intermediate result from $o_1$, instead of a dedicated storage unit, as shown in Fig. 3(b). This example demonstrates that fluid samples can in fact be cached within temporary storage constructed from channel segments. This distributed storage can also overcome the bandwidth limit problem at the ports of the dedicated storage unit illustrated in Fig. 1(c) and Fig. 3(c), where multiple fluid samples must be queued when they access the storage unit simultaneously.

Considering storage minimization during scheduling and channels as distributed storage, we can describe the synthesis problem as follows:

*Inputs*: The sequencing graph of a biochemical assay; the execution times of all operations; the maximum numbers of devices allowed in the chip.

*Outputs*: A schedule minimizing intermediate fluid storage; a channel caching schedule including the locations and the time slots of fluid samples stored temporarily in channels; a compact layout of the biochip.

*Objectives*: Minimizing overall resource usage; maximizing the performance/execution efficiency of the biochip.

## 3 Synthesis of Biochips Considering Storage and Caching

Storage and caching should be considered from scheduling to architectural synthesis to reduce storage requirements and to construct channels for distributed storage. In addition, chip area should be reduced as much as possible by the result of architectural synthesis. Correspondingly, the proposed method includes three major parts: 1) intermediate fluid storage is minimized in scheduling and binding; 2) channel segments that cache intermediate fluid samples are constructed during architectural

**Table 1: Variables and Constraints for Scheduling and Binding**

| Variables: | |
|---|---|
| $O$: | set of nodes (operations) in the sequencing graph; |
| $o_i$: | operation indexed by $i$; |
| $t_i^s$: | starting time of operation $o_i$; |
| $t_i^e$: | ending time of operation $o_i$; |
| $u_i$: | duration of operation $o_i$; |
| $E$: | set of edges in the sequencing graph; |
| $(o_i, o_j)$: | edge from $o_i$ to $o_j$ where $o_i$ is the parent of $o_j$; |
| $u_{i,j}$: | transportation/storage time from $o_i$ to $o_j$. |
| $D$: | set of devices in the biochip; |
| $d_i$: | device indexed by $i$; |
| $s_{i,k}$: | a 0-1 variable representing whether $o_i$ is assigned to $d_k$. |

**Constraints:**

*Uniqueness*:
$$\sum_{k=1}^{|D|} s_{i,k}=1, \quad \forall o_i \in O \quad (1)$$
*Duration*:
$$t_i^s + u_i \leq t_i^e, \quad \forall o_i \in O \quad (2)$$
*Precedence*:
$$t_i^e + u_{i,j} \leq t_j^s, \quad \forall (o_i, o_j) \in E \quad (3)$$
*Non-overlapping*:
$$s_{i,k} + s_{j,k} \leq 1 \text{ if } t_i^s < t_j^e \bigwedge t_i^e > t_j^s, \quad \forall o_i, o_j \in O, d_k \in D \quad (4)$$

synthesis, leading to a distributed channel caching system which fulfills the tasks of transportation and storage at the same time. 3) the planar connection graph from architectural synthesis is refined iteratively to generate a compact physical design.

## 3.1 Minimizing storage in scheduling and binding

Scheduling and binding assign operations in a given sequencing graph to time slots of specific devices. To optimize storage requirements, we formulate the scheduling and binding task as an ILP problem.

The variables and constraints for scheduling and binding are listed in Table 1. The uniqueness constraint (1) specifies that operation $o_i$ should be assigned to one device only once. The duration constraint ensures $o_i$ has enough time to finish. The precedence constraint (3) guarantees that a child operation must be executed later than its parents. Finally, the non-overlapping constraint (4) prevents two operations whose operation periods overlap from being executed by the same device. These constraints are common for high-level synthesis and widely used in synthesis methods for biochips [16].

To minimize the execution time of the assay, another variable $t^E$ is used to represent the latest ending time of all operations, constrained as
$$t_i^e \leq t^E, \quad \forall o_i \in O. \quad (5)$$

By minimizing $t^E$, the operations in the sequencing graph are assigned to proper time slots to produce a compact schedule.

To reduce storage requirements, we introduce an additional *storage minimization objective*. Assume that the pure transportation time from a device to another device is a constant $u_c$. If the schedule produces a transportation time larger than $u_c$, the fluid sample must be cached somewhere before it is used. Figure 4 illustrates the schedules of executing an assay with five operations by two devices. In Fig. 4(b), operation $o_2$ is scheduled before $o_3$. Consequently, the result of $o_2$ must be stored until it is used by $o_4$ and $o_5$, leading to two storage requirements. In Fig. 4(c), $o_3$ is scheduled before $o_2$, leading to only one storage requirement lasting a shorter time. In this example, we can observe that the lifetime of stored fluid samples is determined by the difference between the ending time of the parent operation and the starting time of the child operation $u_{i,j}$ defined in Table 1. Consequently, the total storage requirement can be

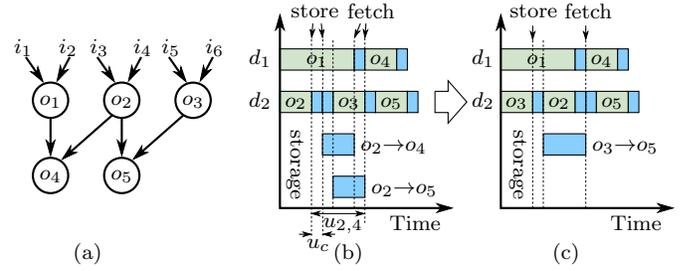

Figure 4: Storage reduction. (a) Sequencing graph. (b) Schedule with two storage requirements. (c) Schedule with one storage requirement. The execution times of the assay with these two schedules are equal.

reduced by solving the following ILP problem
$$\text{minimize} \quad \alpha t^E + \beta \sum_{(o_i, o_j) \in E \bigwedge d_i \neq d_j} u_{i,j} \quad (6)$$
$$\text{subject to} \quad (1)–(5) \quad (7)$$

where $\alpha$ and $\beta$ are constants to control the priority of execution time and storage requirement minimization. $d_i \neq d_j$ excludes the operation pairs assigned to the same device so that they do not need transportation.

## 3.2 Architectural synthesis with distributed channel storage

After solving the optimization problem (6)–(7), we have a schedule similar to Fig. 4(c), including the information: 1) to which devices operations are assigned; 2) the starting time and the ending time of each operation; 3) the starting time and the ending time of each fluid storage requirement.

The schedule, however, only defines the transportation requirements between devices after operations are executed. In the chip, physical channels need be constructed to conduct these transportation tasks. When a large assay is executed by several devices, transportation channels need to be built between nearly any pair of devices to move fluid samples efficiently. Since the flow-layer in a biochip is only two-dimensional, eventually intersections between channels cannot be avoided. At an intersection, a switch should be built to direct the transportation flow to the target device. A switch is constructed by four valves at an intersection, as shown in Fig. 5(a). At a given moment, two out of the four valves in a switch are opened to connect two channel segments. Consequently, a transportation channel between two devices becomes a path formed by several channel segments connected by switches. Such a path is called a *transportation path* henceforth.

Besides channels, the locations of devices should also be determined. These locations should be assigned together with the construction of transportation channels, because the distance and relative locations of devices affect how channels are constructed and how they intersect with each other.

Considering devices and channels together, the architecture of a biochip can be described as devices surrounded by channel segments in the form of a grid. For example, the architecture of a biochip with five devices is shown in Fig. 5(b), where the smaller nodes represent switches and the larger nodes represent devices. Transportation paths between devices are formed from channel segments connected by switches, e.g., path 1 and path 2 in Fig. 5(b). Since transportation paths are used only when there is a fluid sample traveling along it, channel segments can be reused by different paths so that the efficiency of channel segments increases.

With transportation paths formed from channel segments, the proposed distributed channel storage concept can thus be formulated. As illustrated in Fig. 1(c) and Fig. 3(c), a dedicated storage unit suffers bandwidth problem at its ports because mul-tiple fluid samples must be queued when they access the storage unit at the same time. Observing that the storage cells inside a dedicated storage unit are in fact channel segments, as shown in Fig. 1(c), we distribute fluid storage directly in channel segments. For example, in Fig. 5(c), along path 3 a fluid sample is moved to the channel segment between A and B. However, the next operation using this fluid sample is scheduled later, so that the fluid sample must stay in the channel segment. During the lifetime of this storage, the channel segment between A and B cannot be used by other paths and the valves at the two ends of this channel segment should be closed. Further transportation tasks between devices are, however, still fulfilled by paths not including this channel segment, as path 4 and 5 in Fig. 5(d). When the stored fluid sample is finally needed, it is moved to the target device again by a newly constructed transportation path, shown as path 6 in Fig. 5(e).

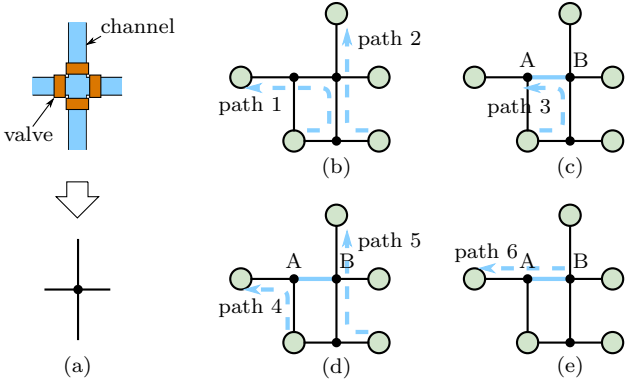

Figure 5: Switch and channel storage. Large nodes represent devices and small nodes represent switches. (a) Switch structure. (b) Two paths sharing one channel segment with time multiplexing. (c) Fluid sample to channel storage. (d) Storage in channel segment. (d) Fluid sample from channel storage to device.

Unlike the dedicated storage unit shown in Fig. 1(c), the distributed storage in a channel segment has a higher access efficiency. When a fluid sample stays in a channel segment, that segment is turned into a *storage segment*. When the fluid sample moves again, the segment becomes a part of the transportation path. This concept of channel role switching unifies transportation and storage, and the low-efficiency channels forming storage cells in a dedicated storage unit are excluded completely. In addition, this on-the-spot caching is closer to the target device than a dedicated storage unit, so that the execution efficiency of the assay can also be improved.

To synthesize the architecture of a biochip from its schedule requires to determine the relative locations of devices as well as channel segments and the switches connecting them as shown by the examples in Fig. 5(b)–(e). The devices, switches and their connections together are called *connection graph*. A valid connection graph should be capable of constructing all transportation paths specified in the schedule and caching intermediate fluid samples in channel segments. To reduce resource usage, the synthesized connection graph should contain as few edges as possible.

The connection graph is generated using a general *connection grid*, as shown in Fig. 6. At each node $n_i$ on the grid, either a device or a switch can be assigned. An edge $e_j$ represents a channel segment capable of caching a fluid sample. We use a 0-1 variable $a_{i,k}$ to represent whether device $d_k$ is assigned to node $n_i$. Since a node can be occupied by no more than one device and a device must be assigned once, $a_{i,k}$ can be constrained as

$$\sum_{d_k \in D} a_{i,k} \leq 1, \quad \forall n_i \in N, \qquad \sum_{n_i \in N} a_{i,k} = 1, \quad \forall d_k \in D \quad (8)$$

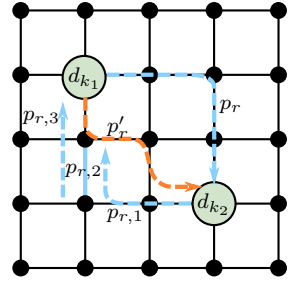

Figure 6: Connection grid.

where $N$ is the set of nodes in the connection grid and $D$ is the set of devices.

Assume there is a transportation path $p_r$ between device $d_{k_1}$ and device $d_{k_2}$ in the period between $t_r^s$ and $t_r^e$, where $r$ is the index of the path. $t_r^s$ and $t_r^e$ are constants determined in the scheduling and binding step in Section 3.1. We use a 0-1 variable $\epsilon_{j,r}$ to represent whether the edge $e_j$ is on the path $p_r$. To construct a path between $d_{k_1}$ and $d_{k_2}$, we need to guarantee that the path starts from the node for $d_{k_1}$ and ends at the node for $d_{k_2}$. Consequently, at one of these two nodes, only one of the four edges incident to the node can be covered in the path. At each other node on the path, exactly two edges are covered by the path, as illustrated with path $p_r$ in Fig. 6. Accordingly, we can construct the path with the following constraints

$$\sum_{e_j \in E_i} \epsilon_{j,r} \geq 2 - a_{i,k_1} - a_{i,k_2} - (1-y_{i,r})M, \qquad \sum_{e_j \in E_i} \epsilon_{j,r} \leq y_{i,r} M \quad (9)$$

where $E_i$ is the set of edges incident to node $n_i$; $y_{i,r}$ is an auxiliary 0-1 variable to indicate whether there is an edge on $p_r$ that is incident to $n_i$; $M$ is a very large constant to transform the two situations indicated by $y_{i,r}$ into linear constraints [17].

The path construction constraints become more complex when a storage is involved, which leads to three sub-paths: 1) from $d_{k_1}$ to a storage segment; 2) the segment caching the fluid sample; 3) from the storage segment to the target device $d_{k_2}$. We denote the three transportation paths as $p_{r,1}$, $p_{r,2}$ and $p_{r,3}$, as illustrated in Fig. 6. Since the end node of $p_{r,1}$ and the starting node of $p_{r,3}$ can be any node on the connection grid as long as they are the two end nodes of the same edge, we use 0-1 variables $a_{i_1,r_1}$ to represent that node $n_{i_1}$ is the last node on the path $p_{r,1}$ and the variables $a_{i_2,r_2}$ to represent that node $n_{i_2}$ is the first node on the path $p_{r,2}$. Afterwards, we create constraints similar to (9) for each sub path. In addition, we include the constraint that $n_{i_1}$ and $n_{i_2}$ are the two end nodes of the same edge.

In the schedule, there are many transportation paths at a given moment. These paths should not intersect at a node or share the same edge to avoid contamination. Therefore, we examine the paths on the connection grid at the starting time of each transportation path, because this is the moment a new transportation is initiated. At each of such moments $t$, we constrain that all the paths existing on the connection grid should not share any edge or intersect at a node, as,

$$\sum_{e_j \in E} \epsilon_{j,r} \leq 1, \quad \forall p_r \in P_t, \qquad \sum_{n_i \in N} \eta_{i,r} \leq 1, \quad \forall p_r \in P_t \quad (10)$$

where $P_t$ is the set of paths existing at time $t$; $\eta_{i,r}$ represents whether node $n_i$ is on path $p_r$; $E$ and $N$ are the sets of edges and nodes in the connection grid, respectively. Exception of constraint (10) is that the two ending nodes of the storage segment $p_{r,2}$ can be passed by other paths when the fluid sample is stored, as path $p'_r$ in Fig. 6, so that their variables $\eta_{i,r}$ are not included in (10) when $p_{r,2}$ exists.

When generating the architecture of the chip, we minimize the number of edges that are really used by the transportation paths to reduce resource usage. If an edge is used once by any transportation path, it should appear in the chip. We use a 0-1 variable $s_j$ to represent whether a channel segment should be kept in the chip, and constrain it as

$$s_j \geq \epsilon_{i,r}, \quad \forall p_r \in P \quad (11)$$

where $P$ is the set including all transportation paths and sub paths.

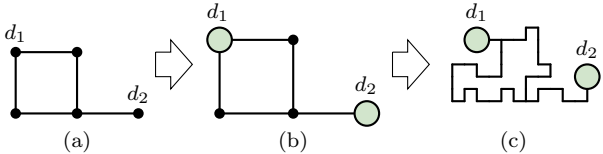

**Figure 7:** Iterative layout compression. (a) Device assignment. (b) Layout expansion. (c) Layout compression.

Finally, the architecture of the biochip can be determined by solving the following optimization problem

$$\text{minimize} \sum_{e_j \in E} s_j \quad (12)$$

$$\text{subject to} \quad (8)\text{--}(11) \quad (13)$$

After determining $s_j$, the edges and nodes that are not used in the connection grid are removed to generate the chip architecture as a planar connection graph similar to Fig. 5(b)–(e).

### 3.3 Iterative compression of chip area

The chip architecture defines the relative locations of devices and their connections by switches. The length of an edge, if used as storage, should be large enough to accommodate one fluid sample. Although we can directly build a biochip with a distributed channel storage architecture as shown in Fig. 7(a)→(b), the size of the chip may be too large because there is much unused space between the channel segments.

To generate the physical design of a biochip from its architecture, we first insert the devices to the connection graph. Since devices are relatively large, we extend the lengths of channel segments to meet all length requirements for channel storage. Afterwards, we collapse the segments toward the upper right corner iteratively. In each iteration, we either reduce the horizontal dimension or the vertical dimension by a constant number and generate bending points on the segments to extend segment lengths. The iterations stop when the layout cannot be compressed further and the result is the final physical design. The major steps of this process are illustrated in Fig. 7.

## 4 Experimental Results

The proposed method was implemented in C++ and tested using a 3.20 GHz CPU with 8 GB memory. We demonstrate the results using three real assays and three randomly generated assays. The information of these test cases are shown in Table 2, where CPA (Colorimetric Protein Assay), IVD (In-Vitro Diagnostics) and PCR (Polymerase Chain Reaction) are real-world assays and the other three assays are randomly generated. The column $|O|$ in Table 2 shows the number of operations in each assay. In the experiments, we used Gurobi [18] to solve the optimization problems.

The result of scheduling with storage minimization is shown in the columns $t^E$ and $t_s$, where $\boldsymbol{t^E}$ is the execution time of the assay defined in (5). The runtime of solving the optimization problem (6)–(7) is shown in the column $\boldsymbol{t_s}$, which was limited to 30 minutes for the solver to return the best-effort results. In architectural synthesis, we use a connection grid to determine device locations and channel segments. The size of the grid is shown in column $\boldsymbol{G}$ in Table 2. After architectural synthesis, the numbers of edges (channel segments) and valves in the chip architectures are shown in the columns $\boldsymbol{n_e}$ and $\boldsymbol{n_v}$, re-spectively. Note the valves counted in the experiments did not include those built in mixers. The runtimes to generate chip architectures are shown in the column $\boldsymbol{t_r}$. In the iterative physical design step, the result of architectural synthesis was first scaled with the unit equal to the minimum channel distance specified by the designer. Afterwards, devices were inserted to nodes determined by architectural synthesis and the layout was iteratively compressed to reduce chip area. The physical dimensions of the chip after architectural synthesis, after device insertion and after iterative compression are shown in the columns $\boldsymbol{d_r}$, $\boldsymbol{d_e}$ and $\boldsymbol{d_p}$, respectively. Since the output of architectural synthesis was already planar, the proposed iterative layout compression can reduce the chip area effectively. The runtime of the physical design phase is shown in the $\boldsymbol{t_p}$ column, which is also acceptable for layout generation.

**Table 2: Results of Scheduling and Synthesis**

| Assay | \|O\| | Scheduling | | Arch. Syn. | | | | Phys. Des. | | | |
|---|---|---|---|---|---|---|---|---|---|---|---|
| | | $t^E$ | $t_s(s)$ | G | $n_e$ | $n_v$ | $t_r(s)$ | $d_r$ | $d_e$ | $d_p$ | $t_p(s)$ |
| RA100 | 100 | 1820 | 1800 | 5×5 | 32 | 58 | 1867 | 20×20 | 26×26 | 16×16 | 68.88 |
| RA70 | 70 | 1180 | 1800 | 4×4 | 20 | 38 | 1819 | 15×15 | 21×21 | 11×12 | 25.37 |
| CPA | 55 | 1070 | 1800 | 4×4 | 20 | 40 | 1817 | 15×15 | 21×21 | 11×13 | 18.45 |
| RA30 | 30 | 670 | 300 | 4×4 | 8 | 16 | 1800 | 15×10 | 21×16 | 13×9 | 0.13 |
| IVD | 12 | 280 | <1 | 4×4 | 5 | 10 | 25 | 10×5 | 16×9 | 12×5 | 0.03 |
| PCR | 7 | 290 | <1 | 4×4 | 5 | 8 | 20 | 5×10 | 7×14 | 4×8 | 0.01 |

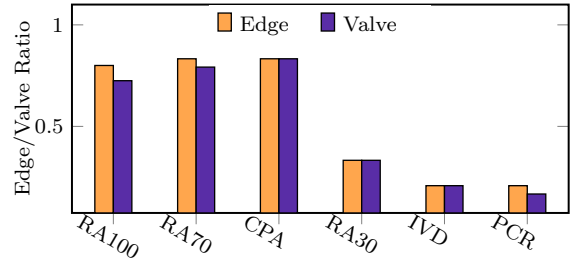

**Figure 8:** Edge and valve ratios in the result of architectural synthesis compared with the original edges and valves in the connection grid.

In architectural synthesis, we start with a connection grid. After synthesis, only the edges that are used at least once are kept in the result. The ratios of the number of used edges to the total number of all the edges in the grid is shown in Fig. 8, where all these ratios are smaller than 1, and a half of them are even close to 0, showing that the architectural synthesis approach confines resource usage effectively on only a part of edges to reduce resource usage. After removing the unused edges, the number of valves is also reduced, as shown by the valve ratios in Fig. 8.

To evaluate the effectiveness of the proposed storage reduction in scheduling and architectural synthesis, we tested the scheduling of minimizing the execution time of the assay only. Afterwards, we applied architectural synthesis to the resulting schedules. Fig. 9 shows the comparison of execution time of assays, the number of edges and the number of valves in the two cases with and without storage optimization. In this comparison, it can be observed that storage optimization generated comparable execution time in the cases IVD and PCR, but the execution time of RA30 is slightly larger, which is acceptable for most biochemical experiments. However, the numbers of edges and valves in the result of RA30 are much smaller, because storage optimization improves the efficiency of channels and thus valves effectively so that fewer resources are required to execute the assays.

In previous methods, the storage and caching problem has not been considered. When there is a storage requirement, it is usually assumed that the intermediate fluid sample is trans-ported to a dedicated storage unit. To compare the efficiency of the synthesized chip architecture with that from the assumed storage transportation, we examine the storage requirement in the schedules generated by the proposed method. When storage requirements appear, they are assumed to queue at the entrance of a dedicated storage unit. The maximum size of the storage cells is the maximum number of fluid samples stored simultaneously in the storage unit. Due to the bandwidth limit of the storage unit, the execution of the assay was thus prolonged. In addition, the dedicated storage unit used many valves to control the access of storage cells. The comparison of channel segments and valves in the result from the proposed method and the result with a dedicated storage unit is shown in Fig. 10. From this comparison, we can see that the execution time and the number of valves are well below 1, leading to a more efficient execution of the assay with fewer resources. For example, the execution time reduction for RA100 has already reached about 28%.

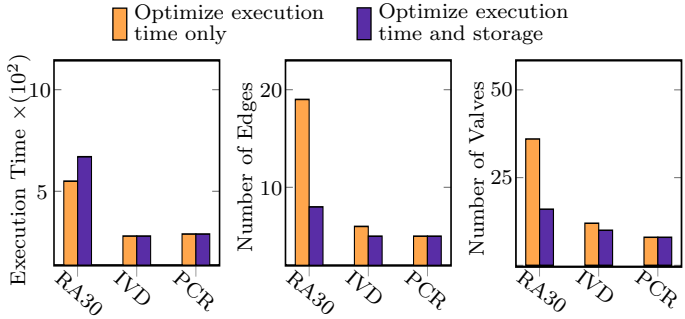

**Figure 9: Comparison of the results with and without storage optimization.**

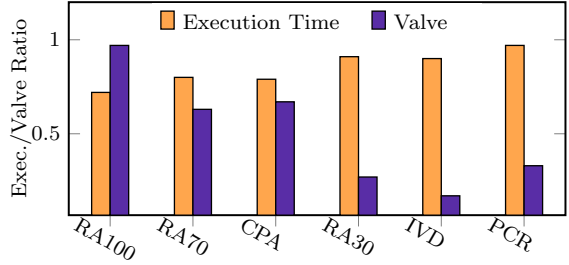

**Figure 10: Comparison of execution time and the number of valves in the results with channel caching and dedicated storage unit.**

Finally, we show two execution snapshots of the assay RA30 in Fig. 11. In Fig. 11(a), a transportation path is formed as $d_2 \rightarrow A \rightarrow B \rightarrow C \rightarrow D$ to store a fluid sample into the channel segment between $C$ and $D$. In Fig. 11(b), a transportation path is constructed as $d_1 \rightarrow D \rightarrow A \rightarrow d_2$ while the channel segment between $C$ and $D$ is caching a fluid sample.

## 5 Conclusion

In this paper, we have proposed the first method to generate a biochip architecture considering storage requirements. By caching fluid samples on-the-spot, transportation channels and storage channels were unified into a connection graph. In addition, transportation paths were constructed on a connection grid dynamically and transportation conflicts were modeled directly instead of being dealt with in a post-processing step. Experimental results confirmed that with this uniform model the architecture generated by the proposed method is more efficient in executing biochemical assays even with fewer resources.

## Acknowledgment

The work of B. Li and U. Schlichtmann was supported by the IGSSE Project FLUIDA of Technical University of Munich.

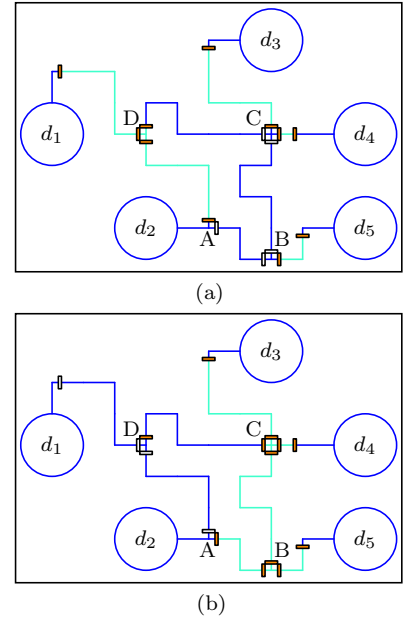

**Figure 11: Snapshots of the synthesized chip executing RA30 at (a) 35s and (b) 45s. The channel segments in blue are transporting or storing fluid samples.**

The work of Chunfeng Liu was supported fully, and the work of T.-Y. Ho was supported in part, by the Technical University of Munich – Institute for Advanced Study, funded by the German Excellence Initiative and the European Union Seventh Framework Programme under grant agreement n° 291763.